# Graphical Abstract

**Acoustic Resonance Effects and Cavitation in SAW Aerosol Generation**

Mehrzad Roudini, Juan Manuel Roselló, Ofer Manor, Claus-Dieter Ohl, Andreas Winkler

# Highlights

**Acoustic Resonance Effects and Cavitation in SAW Aerosol Generation**

Mehrzad Roudini, Juan Manuel Rosselló, Ofer Manor, Claus-Dieter Ohl, Andreas Winkler

- Understanding the nonlinear acousto-hydrodynamics of standing surface acoustic wave (sSAW) interactions with a micro-scale liquid film.

- Demonstration of the presence of micro-cavitation in an MHz-frequency SAW aerosol generator.

- Large volume oscillation from the micro cavitation bubbles in the acoustically induced liquid film is analytically explained.

- The experimental explanation of the droplet generation mechanisms in SAW nebulizers are provided.

# Acoustic Resonance Effects and Cavitation in SAW Aerosol Generation

Mehrzad Roudini[a], Juan Manuel Rosselló[b], Ofer Manor[c], Claus-Dieter Ohl[b], Andreas Winkler[a]

[a]*SAWLab Saxony, Leibniz Institute for Solid State and Materials Research Dresden, Helmholtzstr. 20, Dresden, 01069, , Germany*
[b]*Otto von Guerricke University, Institute for Physics, Universitätsplatz. 2, Magdeburg, 39106, , Germany*
[c]*Technion—Israel Institute of Technology, Department of Chemical Engineering, Haifa, 3200003, , Israel*

## Abstract

The interaction of surface acoustic waves (SAWs) with liquids enables the production of aerosols with adjustable droplet sizes in the micrometer range expelled from a very compact source. Understanding the nonlinear acousto-hydrodynamics of SAWs with a regulated micro-scale liquid film is essential for acousto-microfluidics platforms, particularly aerosol generators. In this study, we demonstrate the presence of micro-cavitation in an MHz-frequency SAW aerosol generation platform, which is touted as a leap in aerosol technology with versatile application fields including biomolecule inhalation therapy, micro-chromatography and spectroscopy, olfactory displays, and material deposition. Using analysis methods with high temporal and spatial resolution, we demonstrate that SAWs stabilize spatially arranged liquid micro-domes atop the generator's surface. Our experiments show that these liquid domes become acoustic resonators with highly fluctuating pressure amplitudes that can even nucleate cavitation bubbles, as supported by analytical modeling. The observed fragmentation of liquid domes indicates the participation of three droplet generation mechanisms, including cavitation and capillary-wave instabilities. During aerosol generation, the cavitation bubbles contribute to the ejection of droplets from the liquid domes and also explain observed microstructural damage patterns on the chip surface eventually caused by cavitation-based erosion.

*Keywords:* surface acoustic wave, acoustofluidics, nebulization,

# 1. Introduction

Compact surface acoustic wave (SAW) aerosol generators, also known as SAW nebulizers or atomizers, are attracting widespread interest in many technical processes due to their capability to produce directed fine aerosols with adjustable narrow size distribution and low shear forces, making them compatible even with complex biomolecules. They show considerable advantages over other traditional aerosol generators and, in particular, ultrasonic atomization. SAW aerosol generators use significantly less power, do not require moving parts, orifices (e.g., nozzles), or meshes, and are suited to produce micrometer-sized droplets with adjustable mean diameters between approximately 0.5 and 30 µm. They support standing or traveling SAWs of a few to several hundred MHz frequencies and with pico- to nanometer mechanical displacement amplitude at the piezoelectric chip surface. The generators are typically constructed from a single-crystal piezoelectric substrate [1], which is patterned by the interdigital transducer (IDT), i.e., a pair of interlocking, comb-shaped metal electrodes [2]. The interaction of SAW with a liquid film placed in their propagation path, a principle first demonstrated by Kurosawa et al.[3], results in complex fluid dynamics, and eventually aerosol generation at increased SAW amplitude. This technology was, however, just recently taken to an application-relevant level [4, 5]. Adjusting the aerosol drop size range is a crucial aspect for various applications, including miniaturized inhalation therapy [6, 7], material deposition [8, 9], liquid chromatography/spectroscopy [10, 11], and olfactory displays [12]. However, we show that the large pressure levels (relative to the low-pressure levels used to actuate liquid in SAW microfluids) further support cavitation. For the first time, we experimentally demonstrate the appearance of cavitation bubbles in the liquid film under SAW load. The bubbles implosion damages the surface of the piezoelectric aerosol generator, leaving erosion which, by itself, is an indication of cavitation; the chipped fragments leave the generator surface via the aerosol. In this paper, we use theory and experiment to study the physical origin of cavitation and additional acoustofluidic mechanisms in SAW aerosol generation in an attempt to provide comprehensive insights for the future prevention of cavitation influence on applications. The mechanism of SAW-induced acoustowetting and, further, aerosol generation comprises influences of acoustic radiation pressure and both the Eckart [13] and Rayleigh streaming [14, 15] mechanisms, as well as other effects, including the liquid geometry [16], wetting properties of the parent liquid [17], and droplet generation on the atomizing platform [16, 18, 19]. However, the SAW atomization mechanism and observed droplet size origin for highly resonant acoustofluidic systems were shown to differ from those described in the available literature [18]. Conventional ultrasonic nebulizers, which generally operate at a few hundred kHz frequencies and require higher electrical power, have been the subject of two main theories and one conjunctive theory despite major variations in the nature and physics of the governing acoustofluidic effects. In the initial hypotheses by Lang [20], the original average size of generated droplets could eventually be estimated based on Kelvin's equation [21], Rayleigh's instability [22] of liquid column or film and the stability limit wavelength. A further hypothesis is the cavitation theory explained the occurrence of droplets as a result of the collapsing of bubbles in the vicinity of the liquid surface [23, 24]. The theory of a cavitation-induced atomization mechanism under MHz ultrasound irradiation was sometimes rejected due to the high threshold power for cavitation in this frequency range. Bograslavski and Eknadiosyants et al. [25] proposed the conjunction theory, where both mechanisms are interrelated.

This paper investigates the complex acousto-hydrodynamics associated with standing surface acoustic wave (sSAW) interactions with a liquid film, its acoustic stabilization, patterning, and the droplet breakup mechanisms from the developed liquid pattern by employing analysis techniques with unprecedented high temporal and spatial resolution. Experimentally shown and supported by analytic models are the dynamics of the cyclic, spatiotemporal fluid domes formation under standing SAW and their transformation into acoustic resonators with oscillating pressure amplitudes in which cavitation bubbles may be nucleated. For the first time, the observed fragmentation of liquid domes can be linked to three separate droplet generation mechanisms, including cavitation and capillary-wave instabilities. Interestingly, the locations of recorded bubble oscillations coincided with observed microstructural damage patterns, revealing cavitation-based erosion mechanics in SAW nebulization. This work contributes to a deeper understanding of acoustic resonance effects in SAW aerosol generation, enhancing the applicability of such devices in real-world applications.

## 2. Material and methods

### 2.1. SAW microfluidic chip

This study uses standing surface acoustic waves (sSAWs) to disintegrate an acoustically stabilized liquid film with a thickness in the order of the SAW wavelength into fine droplets. Partially open microfluidic channels and interdigital transducers are structured on a 4" single-side polished lithium niobate wafer substrate with X-propagation direction (128°YX LiNbO3), subsequently diced into single SAW aerosol-generation chips ($8 \times 19\,\text{mm}^2$). In the current chip layout, two interdigital transducers (IDT) ($\lambda/4$ type, $90\,\mu\text{m}$ wavelength, 46 electrode pairs, $0.5\,\text{mm}$ aperture, matched to $50\,\Omega$ impedance by the number of finger electrodes) are opposing each other with a distance of $6\,\text{mm}$ for sSAW excitation based on superposition of two counter-propagating traveling SAWs. The double-side polished, transparent LiNbO3 was used for the fabrication of SAW-chips to access optically the atomization zone from below through the chip substrate. Due to the birefringent nature of the substrate, double images are recorded in optical transmission when no polarization filter is used, as was done in this study to not reduce exposure. SU-8-50 photoresist (Microchem Corp.) was spin-coated on each chip surface. The SU-8 was structured via double-exposure photolithography to form chip-integrated microchannels. The construction and configuration of the SAW chip were described earlier [4, 5]. A microscopic image of a microfluidic SAW atomization chip with its components is shown in Figure 1a.

### 2.2. SAW nebulizer

Flat, anodized aluminum chip holders with two adjustable PCB carriers are used as chip holders in the current experiment. A chip holder with an integrated hole was prepared to access the atomization zone optically from below through the transparent chip substrate, as shown in Figure 1b. The liquid was introduced from a syringe pump to the microchannels via sealing O-rings, a 90-degree bent syringe needle, silicon tubing, and PTFE tubing. The customized printed circuit board (PCB) comprises strip lines matched to $50\,\Omega$ impedance, female SMA PCB connectors, and gold-coated spring pins, which directly contact the pads on the chip surface. Signals were supplied at the operating frequency of the IDTs via SMA cables from a dual-channel signal source (BSG F20, BelektroniG GmbH, Germany).

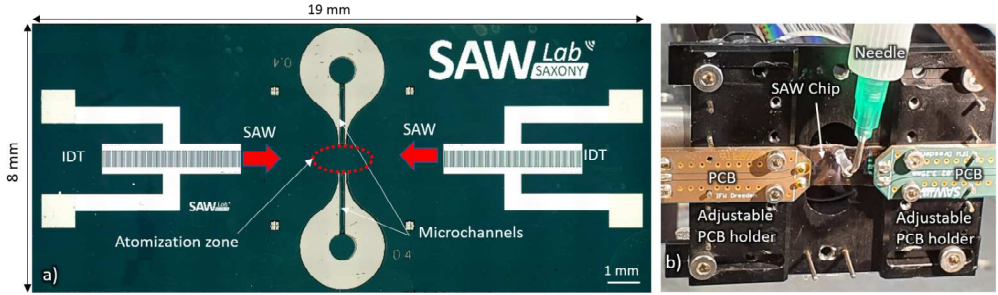

Figure 1: a) Microscopy image of SAW chip for aerosol generation includes IDTs with a 90 µm wavelength and integrated microchannels located outside the acoustic aperture, i.e., in the boundary region of the acoustic beam. b) The assembled chip holder and its components to visualize the atomization zone optically from below through the transparent chip substrate.

*2.3. Optical imaging setup*

The formation process of the liquid domes and the rather fast bubble dynamics were captured using high-speed photography combined with a long-distance optical microscope, as illustrated in Figure 2.

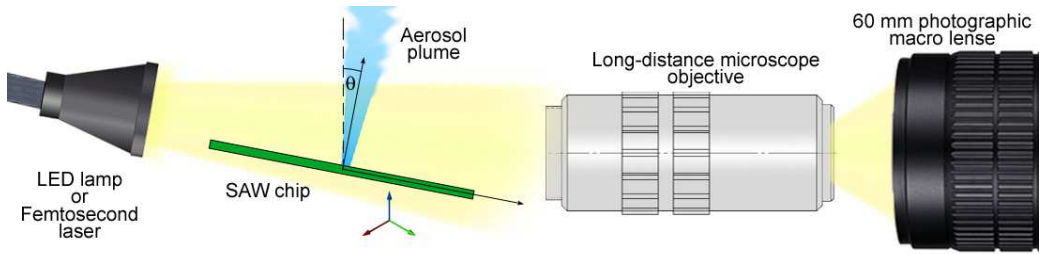

Figure 2: Overview of the optical setup used in the experiments. The SAW chip could be rotated to observe the drop formation and the atomization process from different perspectives.

The camera (XPV-X2, Shimadzu) could reach frame rates up to 5 Mfps (i.e. 200 ns between frames). The illumination was selectively performed using and white LED lamp (SMETec with 9000 lm) or a femtosecond laser (FemtoLux 3, Ekspla, $\lambda =515$ nm, pulse duration of 230 fs). While the LED light allowed us to observe the cavitation bubbles in the interior of the drop, the ultra-fast pulses of the laser illumination were more suitable for capturing sharp images of the tiny aerosol particles, i.e., without motion-induced blurring. A uniform laser illumination was achieved by transporting the light beam with a 600 µm optical fiber and then expanding it with a coupling lens (F230SMA-A, Thorlabs). The microscopic images were taken using a long-distance microscope objective (Edmund Optics) with different magnification factors (e.g. 5X, 20X, or 50X) complemented with a second macro lens (f2.8 macro lens, LAOWA). The position of the SAW chip could be changed with a micrometric XYZ stage and the observation angle ($\theta$) could be adjusted by changing the chip orientation, as shown in Fig. 2. An inclination of the SAW chip plate of around 10was appropriate for visualizing both the drop shape and the aerosol particle ejection. Alternatively, we could rotate the plate 90to see the droplet's interior and the bubble formation process in greater detail, e.g., see Fig. 4.

## 3. Results and discussion

*3.1. Stability and instability in the liquid film*

In this study, the liquid atomization zone on the piezoelectric substrate, i.e., the zone of SAW-fluid interaction and aerosol origin, is optically visualized using a custom SAW aerosol generation chip design with maximum process control in an ultra-high-speed camera to gain more insight into the hydrodynamics associated with standing SAW-liquid interaction. On the surface of the microfluidic chip, Figure 3a depicts a schematic of the atomization zone in front of the microchannel. While other physical effects like acoustowetting are occurring in this zone, the most important for the aerosol generation is the formation of fluid domes [18]: Figure 3b shows a microscopy image of the atomization zone, including three important regions explained in more detail below. As shown in Figure 3c (recorded at a frame rate of 2 million fps), sSAW-induced quasi-stable liquid dome-shaped patterns are periodically observed at a distance of half of the SAW wavelength $\lambda_{SAW}/2$ at the location with a sufficiently high SAW amplitude during liquid atomization. This result highlights the importance of the acoustic wave field profile in forming the liquid film pattern in the atomization zone. Moreover, the results of our current experiments found clear support for the dome-shaped patterns being subject to cyclic, spatiotemporal changes with a typical time constant of a few microseconds, and being the sole origin of the generated droplets during the sSAW-fluid film interaction, respectively.

The recorded time series of microscopic images in Figure 3d show typical cyclic variation in the shape of the liquid patterns induced by sSAW, observed in this study for the first time. A description of the experimental setup is provided in the Methods section, and a video capturing the phenomena is included in the Supplementary Information. Suppose a liquid film exists in a SAW field region with a sufficiently high displacement amplitude. In that case, the liquid film's local volume can eventually reach a critical value in terms of its surface energy [16]. As a consequence, liquid droplet-shaped patterns are formed as a consequence of a standing longitudinal pressure field in the SAW antinode regions and are internally dominated by Eckart streaming. The droplet-shaped patterns are probably consistent with the observed translating solitary wave-like structures caused by the leaked acoustic energy from a traveling SAW in a previous study on oil films under SAW influence [17]. While the formation of the resonant longitudinal pressure fields can lead to this dome formation in different SAW types (standing and traveling), the constant location of the observed patterns in this study, is contrary to their Eckart-streaming induced directional drifting under traveling SAW influence. The whole event is shown in video S1 in the Supplemental Materials. The growing dome-shaped patterns are continuously supported with liquid by a very thin and modulated liquid film, shown as region 2 in Figure 3b, from an extended, chaotic, surface fluctuated film in front of the microchannel outlet. The maximum height of the observed domes was around 15 µm, reached after about 18 µs, which is in good agreement with the results of the previous study [18], irrelevant to electrically applied power, introduced liquid flow rate, and SAW type.

*3.2. Bubble formation and collapse*

While investigating aerosol generation using high-speed video recording from a top view, single and multi-bubbles can be observed in different locations within the liquid domes. Cyclic bubble appearance, growth, and collapse events are depicted in Figure 4. We observed, that When the liquid dome height reaches a specific value in the wavefield, as explained in the previous sections, cavitation bubbles are nucleated in the half-side of the dome (shown in Figure 4a – 0.2 µs). The single bubble expands in size, reaching its maximum equivalent diameter of 12 µm in 0.6 µs and collapses after approximately 1.2 µs. At 1.4 µs, the bubble collapses near the free boundary, leading to a liquid jet formation, which is shown by a white arrow in Figure 4a. A similar jetting mechanism was found for bubbles bursting on a free surface [26, 27]. In that study, the bubbles are driven by the buoyancy force towards the free surface and produce a liquid jet that releases liquid particles into the atmosphere when the cavity opens. In the present case, the strong acoustic gradient inside the liquid drop pushes the bubble towards the free surface (i.e., through primary Bjerknes force) and gives origin to the jetting [28]. After the mass ejection, the dome slowly regains a nearly half-spherical shape, and new cavitation bubbles appear, restating the jetting process. Interestingly, the nucleation of cavitation bubbles takes place only after the domes reach a fully developed shape (see Figure 3c). This is a strong indicator of the acoustic resonance nature of the cavitation process.

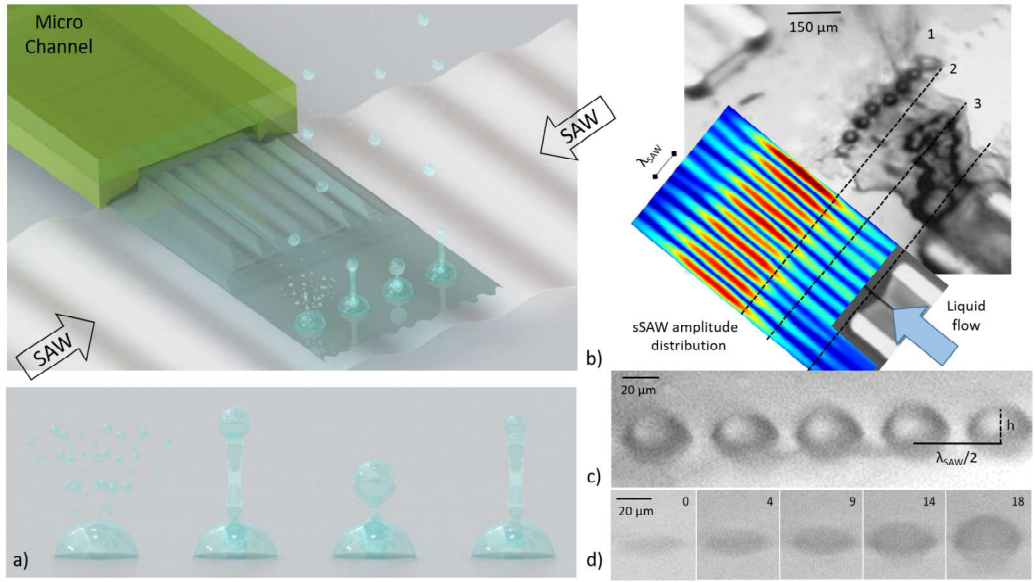

Figure 3: a) Schematic representation of the microacoustic aerosol generation platform, where droplets are formed from acoustically induced fluid domes. b) Microscopy images showing regions and effects within the atomization zone during the aerosol generation in a standing SAW wave field with 90 µm SAW wavelength: The atomization zone in front of the fluid delivery channel can be separated into three regions: zone 1, 2 and 3. The measured sSAW amplitude distribution in front of the microchannel outlet and with respect to the atomization zone. More detailed wavefield measurement setups are explained in Supplemental Material. c) sSAW-induced dome-shaped liquid pattern in the atomization zone appears quasi-static at a lower frame rate (2 million fps). d) Time series images of the cyclic liquid dome growth under periodic SAW influence ($T \approx 22.5 ns$). See Supplemental Material in video S1. Time is in microseconds and the image resolution is 0.6 µm per pixel.

In the following experiments, the atomization zone has been optically investigated from the bottom of the transparent, double-sided polished sSAW chip to obtain more insight into cavitation bubble generation inside the liquid domes during droplet formation. The acoustically induced bubbles frequently appear in the center of domes, similar to the shown event in Figure 4b. Moreover, multi bubbles are occasionally formed, as shown in Figure 4c in 0.2 µs, simultaneously in the liquid pattern. As seen in the recorded microscopy images in Figure 4c, bubble formation, and collapse events last longer for the more giant nucleated bubbles.

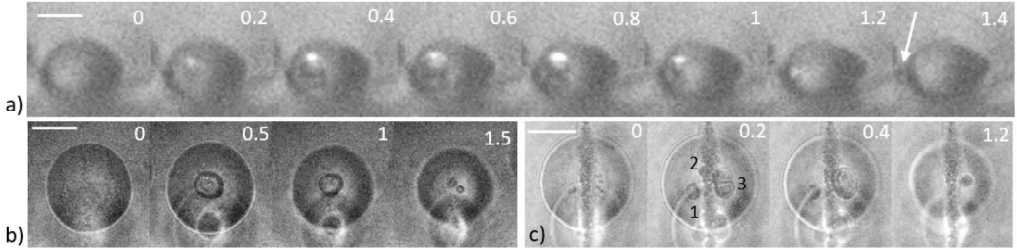

Figure 4: Microscopic observation of bubbles inside the liquid dome and cavitation bubble dynamics. a) A time series from a single bubble's formation and collapse inside a liquid dome during SAW liquid atomization. b,c) Time series from the bottom of the transparent chip showing b) single and c) multi-bubble dynamics in the dome. The length of the bar in the upper left corner of the first frames is 20 µm; Time in µs. The transparent, but birefringent substrate induces a double image of observed droplets in b,c.

*3.3. Theory*

In the following chapters, we explain our observations of the liquid film resonance and droplet breakup in more detail and present a new theory for dome formation and their cyclic, spatiotemporal behavior.

*3.3.1. Origin of domes*

In 3.1 we demonstrate for the first time cyclic, spatiotemporal variations in the shape of the liquid film in the atomization zone, which support cavitation and aerosol generation. In particular, we observe that under a standing SAW at a frequency of 43 MHz in the substrate, a water film appears to take the form of domes periodically at a distance of approximately 45 $\mu$m, equal to half a SAW wavelength, in the maximum amplitude regions. The heights of the domes, appearing quasi-stable at a lower imaging frame rate, are approximately 15 $\mu$m. The domes appear from a liquid film of approximately half their height. The heights of the domes are connected to the ultrasound leakage off the SAW: The standing SAW leaks a longitudinal pressure wave of the same frequency normal to the surface into the fluid. The wavelength of the ultrasonic leakage to the water film in our experiment is approximately $\lambda = 1500\,\text{m/s} / 43\,\text{MHz} \approx 35\,\mu\text{m}$. Hence the thickness of the flat part of the film corresponds to approximately $7.5/35 \approx 0.21$ of the ultrasonic wavelength. The measured height of the flat component of the film supports near anti-resonance of the ultrasonic leakage in the film. The corresponding acoustic radiation pressure in the flat film is near minimal. Moreover, the maxima of the domes correspond to approximately $15/35 \approx 0.43$ of the ultrasonic wavelength. The film's maximum height supports the near resonance of the ultrasonic leakage in the film. The domes appear to be directly connected to the SAW in the solid substrate, both in length and height.

Since the SAW is a standing wave, one may assume that the normal vibration velocity of the SAW is approximately $U\cos(\kappa x)\cos(\omega t)$, where $U$ is the amplitude of the particle velocity at the solid surface – the amplitude of the normal velocity of the surface mass – and $\kappa$ and $\omega$ are the SAW wavenumber and angular frequency, respectively. The ultrasound leakage penetrates into the liquid, whose free surface (air/water interface) is nearly a perfect acoustic reflector.

Accounting for the presence of near resonance effects gives the excess steady pressure in the dome, [29]

$$\langle p - p_0 \rangle \approx (1 + B/2A)\,\langle E \rangle\,(1 + \sin(2\kappa_l h)/2\kappa_l h)\,, \qquad (1)$$

where the slope of the air/water interface is small and where $B$ and $A$ are the Fox and Wallace coefficients for the non-linear connection between pressure and density, $\kappa_l \equiv 2\pi/\lambda_l$ is the wavenumber of the ultrasonic leakage, $h$ is the local height of the water film, $\langle E \rangle \approx \rho U^2 \cos^2(\kappa x)/4\sin^2(\kappa_l h)$ is the local average excess energy in the water film, and $\rho$ is the average water density, which is the water density at rest to leading order. In Figure 5a, we plot the radiation pressure in the film as a function of film thickness, $h$. The shaded parts do not support stable film thickness due to capillary-radiation pressure instability in oil [17, 30, 31] and in water [32, 33]. The experimentally obtained values are within the stable regime.

To show that the acoustic pressure distribution in eq. 1 is responsible for the dome morphology in the experiment, we solve for the balance between the capillary and acoustic pressure in the liquid film [34] using the augmented Young Laplace equation,

$$\frac{\gamma h_{xx}}{(1 + h_x^2)^{3/2}} = \langle p - p_0 \rangle + \Delta P, \qquad (2)$$

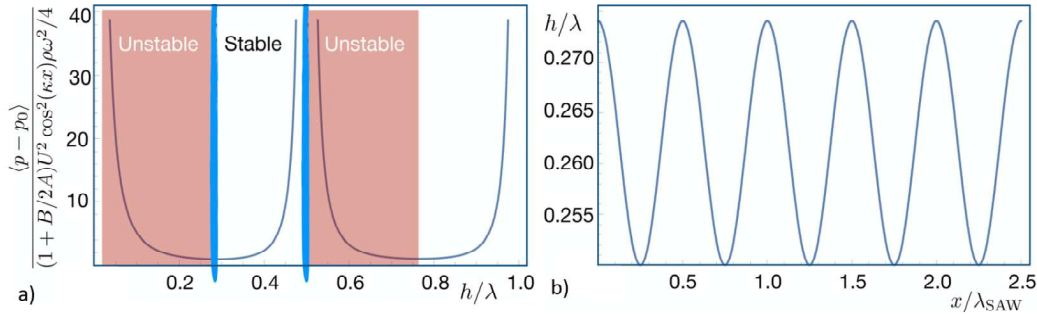

Figure 5: Analytical results for fluid film thickness under SAW influence: a) Film thickness variations under acoustic radiation pressure, where the stable film thickness regime that corresponds to the experiment should be in the approximate range $h \in (h = 8.8-17.5)$ μm. b) Spatial variations of film thickness along the sSAW propagation direction for $(1 + B/2A)U^2 \rho \lambda_{\mathrm{SAW}}^2 / 4\gamma\lambda = 4$.

$$h(x = \lambda_{\mathrm{SAW}}/2) = h(x = 3\lambda_{\mathrm{SAW}}/2) = \lambda/4,$$
$$\frac{\partial h}{\partial x}(x = \lambda_{\mathrm{SAW}}/2) = 0.$$

The equation and boundary conditions give the geometry of the free film surface, where $\Delta P$ is a constant excess pressure in the film to satisfy mass conservation. The boundary conditions restrict the thickness of the film between domes to $h = \lambda/4 \approx 8.8$ μm and require that it is flat. The result in Figure 5b portrays an array of domes that naturally appear from the solution when integrating the problem in $x \in (0, 5\lambda_{\mathrm{SAW}})$. Hence, the second-order steady component of the acoustic radiation pressure, generated by the SAW, may support the formation of the observed domes. Next, we show that the leading order transient component of the acoustic radiation pressure may support the observed cavitation effects.

*3.3.2. Oscillating pressure in the domes*

Our experiment indicates that the domes support cavitation in their midst. This most likely results from the leading order acoustic pressure fluctuations that the SAW generates in the domes. In particular, the steady, second-order acoustic radiation pressure in eq. (1), which contributes to the shape of the domes in eq. (2), is further accompanied by much stronger, leading-order transient acoustic pressure fluctuations.

The leading order transient component of the acoustic radiation pressure is best described by using the Lagrangian particle displacement (La-grangian strain) transverse to the solid surface, $\zeta_L$, in the domes. In our current problem, it is given by $\zeta_L \approx (U/\omega) \sin(\omega t) \sin(\kappa(h_{\max} - a))/\sin(\kappa h_{\max})$, where $a$ is a Lagrangian vertical coordinate which vanishes ($a = 0$) at the solid surface. The Eulerian position of the solid surface is given by $\zeta_E = (U/\omega) \cos(\kappa x) \sin(\omega t))$, so that $\zeta_L = \zeta_E(x = a, t) - a$. The corresponding leading order pressure amplitude in the domes is given by [29]

$$|p - p_0| \approx \frac{\rho \omega}{\kappa} \left|\frac{\partial \zeta_L}{\partial t}\right| \approx \frac{\rho \omega U}{\kappa} \sin(\kappa(h_{\max} - a))/\sin(\kappa h_{\max}), \qquad (3)$$

where $\rho\omega/\kappa$ is the amplitude of acoustic impedance. We demonstrate position variations of both the Lagrangian amplitude of the particle velocity, $\zeta_L$, and the local gauge pressure amplitude, $|p - p_0|$, in domes of different heights in Figure 6a.

One would expect that cavitation bubbles appear near the middle height of the domes in our experiment, where the oscillating pressure, $p - p_0$, is maximized under the dome heights of $h_{\max} \approx 0.43\lambda$. However, the experiments show that cavitation is more likely to occur at the solid surface, even for pressures below the one found at the antinode. This situation usually takes place when the surface tension of the solid/vapor interface is smaller than the tension between the vapor and the liquid phases [35]. At the same time, the presence of small crevices on the surface can trap gas molecules that could act as nucleation sites when driven by the ultrasound [36]. The maximum normal velocity of the solid surface under the domes is in the range $U = 0.1 - 0.6$ m/s. The amplitude of the acoustic impedance in water, $\rho\omega/\kappa \approx 1.5 \times 10^6$ kg/m$^2$s gives a characteristic oscillating pressure amplitude in the domes in the range of $\rho\omega U/\kappa \approx 0.15 - 0.90 \times 10^6$ Pa or $1.5 - 9.0$ bar. This Figure increases as the maximum heights of the domes approach their singular value at $h_{\max} = 0.5\lambda$ in eq. (3). For example, for $h_{\max} = 0.43\lambda$ we observe a magnification in particle displacement by a bit more than a factor of 2. For $h_{\max} = 0.49\lambda$ we observe a magnification in particle displacement by a factor of 16.

*3.3.3. Nonlinear response of a gas bubble*

The bubble visible in Figure. 4 has a radius of several micrometers. A bubble of this size is far larger than the linear resonance radius; thus, for a harmonic oscillator, one would expect that the bubble oscillates 180 degrees out of phase with the driving pressure.

We speculate that the growth of the droplet to resonance size amplifies the pressure that a bubble will eventually experience, e.g. to 140 bar at $h_{\max} = 0.49\lambda$. Now, we can consider that a small gas bubble forms within the droplet, e.g., through degassing, entrainment, or heating of the surface. Then it will respond to the acoustic pressure within the droplet. Figure 6b depicts such the spherical response for a bubble with an initial bubble radius of $1\,\mu$m in water using the Rayleigh Plesset model that accounts for viscosity and surface tension when driven at 43 MHz. At a low driving pressure of 5 bar the bubble oscillates 180 degrees out of phase with respect to the driving pressure $p(t)$ with a small amplitude approximately harmonically around its rest radius. With increasing pressure, here 50 bar the out-of-phase oscillation remains, yet the mean bubble radius is considerably smaller than its rest radius. The non-linear response of the radial oscillation increases with increasing driving pressure. At $p_A = 100$ bar strong fluctuations in the bubble oscillations are observed. Over time, the bubble is more strongly compressed than expanded and eventually undergoes strong collapses, i.e., at $t = 0.6\,\mu$s and at $t = 0.85\,\mu$s. The inset in Figure. 6b reveals the out-of-phase oscillation of the bubble by comparing the time-varying radius with the pressure, $p(t)$, sketched with a dashed line.

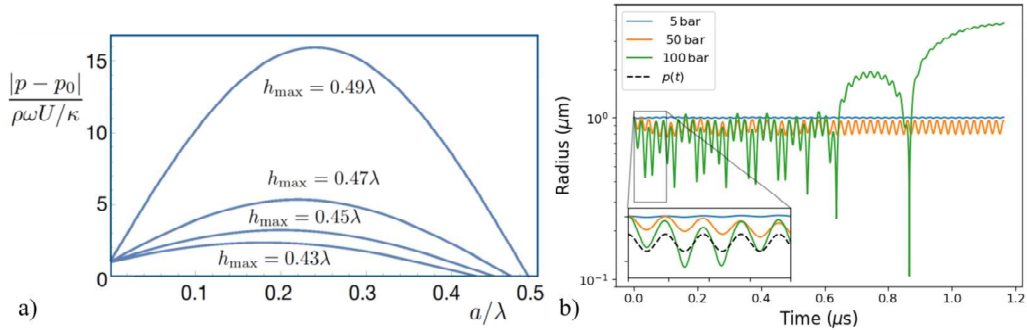

Figure 6: a) Vertical spatial variations of the gauge pressure amplitude along the (vertical) middle axis of the domes for different maximum dome thicknesses, where the maximum particle displacement and hence pressure amplitude appears at mid dome height, $a \approx h_{\max}/2$. b) Response of a $1\,\mu$m bubble driven above resonance frequency for increasing driving amplitudes of 5, 50, and 100 bar. At 100 bars, the bubble undergoes a violent inertial collapse at $t = 0.85\,mu$. In the inset, the driving pressure (not to scale) is plotted as a dashed line demonstrating the out-of-phase response of the bubble oscillations.

While this Rayleigh-Plesset model with its many simplifying assumptions is far from reality, it provides insight into how inertial bubble collapses within the droplet may occur. Here, although the bubble is considerably smaller than its linear resonance size, it can be driven to large expansions by the compression of the non-condensable gas. This compression then results in larger expansions and even stronger collapses. Over time, a threshold compression may be achieved that decouples the gross bubble dynamics from the acoustic driving and results in a violent inertial collapse. While we are not aware that this regime of very strong driving of bubbles far above their linear resonance frequency has been studied, it reminds us of the previously studied non-linear resonance of bubbles in single-bubble sonoluminescence (SBSL). In SBSL the bubble is driven at a frequency much lower than its linear resonance size. There, inertial collapse occurs once the driving pressure overcomes the Laplace pressure [37].

*3.4. Role of cavitation on the liquid atomization*

This section will explain the fragmentation process of acoustically induced liquid patterns and provide details of its dynamics.

Three general liquid aerosol generation mechanisms are observed using high-speed microscopy video recording as shown in Figure 7. In the first observed regime, a cloud of fine droplets of less than 1-2 µm is fragmented from the entire dome surface (shown in Figure 7a and video S2 in Supplemental Material). Since the dome is almost intact during the low-velocity droplet disintegration process, we suggest capillary waves along the surface of the fluid domes to mainly contribute to this generated micro-sized droplet mist. Capillary wave theory indicates that droplet breakup occurs when unstable oscillations disintegrate the crest of capillary waves at the air-liquid interface. High-frequency sSAW transmitted energy into the liquid dome surface to induce capillary waves from which the mist cloud is atomized. Previous experimental studies have similarly shown the appearance of fine droplets in the same size range from a thin liquid film under ultrasonic substrate wave vibrations [16, 38].

A time series image shown in Figure 7b illustrates a typical case of bubble collapse-induced microjet ejection. A microbubble forms, grows, and collapses in the vicinity of the liquid dome interface with air at around 1 µs, shown with a black arrow. After the collapsing phase of the bubble, a spike jet formed at the surface at 1.2 µs as a result of the high-pressure gradient spot between the upper surface of the bubble and the free surface. A microjet with a velocity of about 12 m/s emerges from the dome top surface, and Rayleigh Plateau instability causes the jet to break up into a single droplet with a diameter of 5 µm at 2.8 µs. The whole incident is shown in video S3 in the Supplemental Materials.

The third observed liquid dome fragmentation is a thin liquid film rupture followed by a large droplet pinch-off. Once the liquid dome developed to a certain height, as shown in Figure 7c at 0 µs, we believe a relatively large cavitation bubble formed due to the acoustically induced pressure gradient in the dome. The recorded video frame rate of 500 ns was not fast enough to capture the bubble formation event. However, the next bubble formation event was caught between 11 µs and 18 µs with a similar atomization regime. As a consequence of the abrupt bubble formation and expansion, the liquid film between the bubble and the surrounding air becomes thinner until it eventually ruptures into a cloud of fine droplets at 0.5 µs. The cloud droplets are expelled with a velocity of around 40 m/s from the dome interface. Soon after the rupture of the dome wall, at 2.5 µs, during bubble expansion, the top of the bubble is entrained into the base of the raised free surface, thus causing a substantial disintegration of the droplets from the dome's top surface. The whole event is shown in video S4 in the Supplemental Materials. The stable disintegrated droplet size from the bubble collapse-induced liquid jet is around 20 µm with an approximate initial velocity of 5 m/s.

Based on observed atomization regimes, generated droplets from the atomization zone on an sSAW chip surface consist of two main droplet size classes. One group contributes to the fine droplets with an initial droplet diameter of less than 1-2 µm, mainly disintegrated by the induced capillary waves and cavitation bubble dynamics. A second droplet group includes considerably larger droplets, which are individually broken up by cavitated bubble growth and collapse events. All atomized droplets in an aerosol plume at a few millimeters distance from the chip surface are captured and are clearly presented in Figure 7d. The first group appears in the microscopy image in the form of a cloud of droplets, highlighted with a white arrow, and the second group appears in the form of single droplets, highlighted with black arrows. These observations strongly confirm the measured droplet size distributions using the same sSAW chip layout and DI water in the previous study [18]. In that study, two different droplet-size ranges were indicated as two separate peaks in the measured droplet-size distribution, with all droplet diameters below 30 µm. Notably, the described cavitation-generated droplet ejection differs from the SAW-driven droplet jetting caused by SAW streaming [39].

After evaluating atomization regimes in 200 different captured videos, it appears that the bubble size and the vertical distance from the initial center of the bubble to the free surface are two critical parameters determining the cavitation bubble-induced liquid atomization regimes.

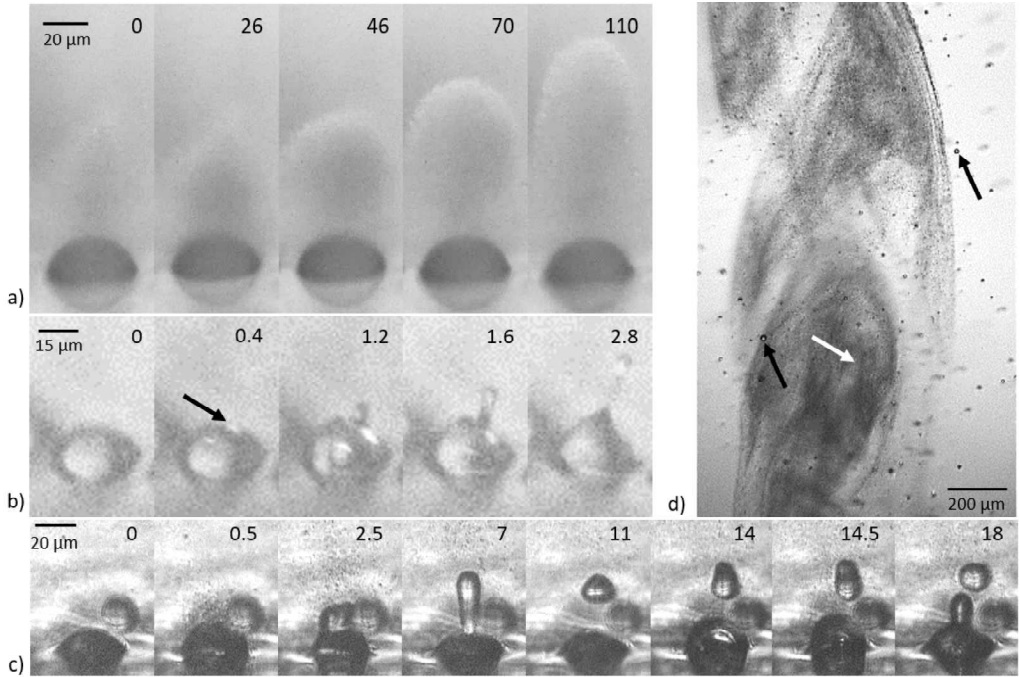

Figure 7: Droplet formation regimes: a) capillary wave-induced mist generation (See Supplemental Material in video S2). b) microjets in the course of the collapse of cavitation bubbles (See Supplemental Material in video S3). c) atomization followed by a large droplet disintegration (See Supplemental Material in video S4). d) Visualization of the ejected droplets during SAW atomization proves two different droplet-size ranges. The time is stated in microseconds.

The results are generally similar to the experimental observation of fragmentation of acoustically levitating droplets by laser-induced cavitation bubbles in [40]. Here, the resonance acoustic-induced pressure distribution in the stabilized dome-shaped pattern at different time scales is attributed to the formation and growth of formed cavities compared to nucleating the cavitation bubble using the laser energy in their study. In future work, it might be important to study how the cavitated bubble moves and where it is, as well as how it relates to the applied sSAW amplitude.

*3.5. Damage on the SAW device*

During the analysis of SAW atomization chips used in our current and previous studies, damage to the substrate surface, i.e., the native LiNbO3 surface and the overlying SiO2 layer, was observed (shown in Figure 8a). The so far unreported damage mechanism leads to the detachment of initially small, later larger particles from the substrate, which can lead to functional impairment or even failure of the device. Furthermore, the detached particles are expected to be part of the aerosol stream, where they could have a significant influence, especially in medical applications like inhalation therapy.

The damage occurs in the regions of high SAW amplitude within the atomization zone following the standing wave antinodes or, if traveling waves are applied, in an extended surface region. In each investigated case, an overlying liquid film was required for the damage, while the other areas of the chip exposed to high SAW amplitudes remained intact. The damaged locations observed in SEM images, as detailed in Figure 8b, seem to be precisely aligned with the observed droplet-shaped liquid patterns where the aerosol droplets are generated. Chemical or electrochemical corrosion can be excluded from the materials used, and the damage timescale is much too small for acoustic or stress migration. The latter mechanisms have been studied intensively [41, 42]. They can be controlled by countermeasures, e.g., improved electrode material systems, surface interface modification [43], and efficient heat dissipation. The new damage mechanism observed seems to be reproducible, time-dependent, and strongly dependent on the liquid's properties, the acoustic amplitude, and non-linear response behavior.

Our experimental evidence supports to claim for the first time that cavitation erosion is the reason for the observed damage spots in the atomization zone on the chip surface. The time series image captured from the bottom of the chip shows the micro cavitation bubble formation location, growing and collapsing in the center of liquid domes and above the eroded regions (Figure 8c). The bubble expands and collapses only in 1.2 µs within an intact dome, followed by the formation of microjets in the collapse of cavitation bubbles at 1.8 µs from the upper dome surface. Two adjacent bubbles likewise appeared in the lower dome at 0.6 µs and collapsed approximately on the same time scale. In addition, these results provide information about cavitation erosion spots with a precise distance of half of the SAW wavelength, during the aerosol generation in the atomization zone. The existence of the observed damage mechanism is highly critical for resonant acoustofluidic devices as it can reduce the device lifetime, alter the device behavior, and cause

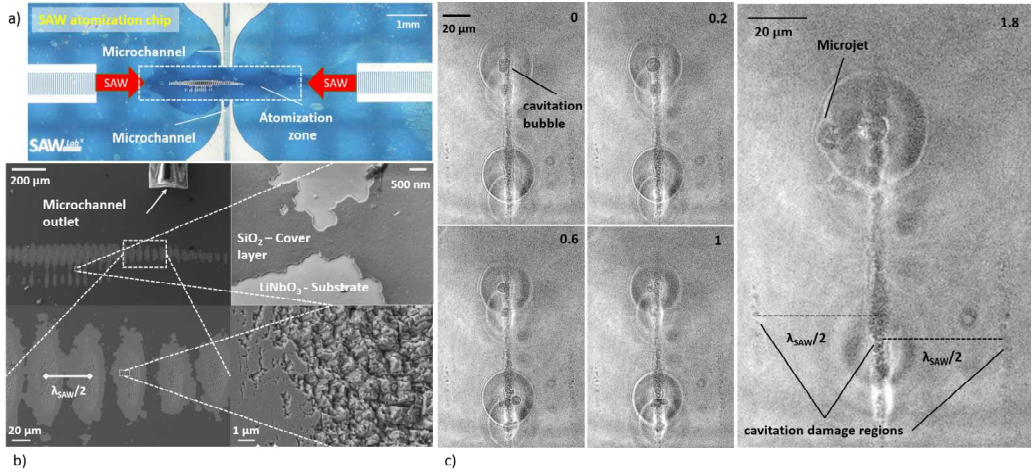

Figure 8: Microstructural damage by cavitation erosion: a) A microscopy image of a used sSAW atomization chip with damage patterns in the regions of high SAW amplitude within the atomization zone. b) SEM images of typical damage pattern in the atomization zone after prolonged exposure to a standing SAW during the liquid atomization on the surface of a LiNbO3 substrate and in a region of a damaged and partially detached SiO2 cover layer on a LiNbO3 substrate with commencing substrate damage. c) The time series images from the bottom of the transparent SAW chip during the atomization process show the exact alignment of the damage pattern in the center of the liquid domes, and bubble formation in a standing SAW wavefield with 90 µm wavelength. Time is in microseconds and the image resolution is 0.68 µm per pixel.

micro- and nanoparticle generation critical for future applications, especially medical devices. On the other hand, acoustic cavitation with violent bubble collapse is often utilized, such as in cancer treatment or extracorporeal shock wave lithotripsy. We would like to emphasize, that the type of SAW operation, e.g. standing or traveling wave, may not be important for the existence of cavitation or cavitation erosion, as both are caused by longitudinal wave (not SAW) resonance, but that it may influence significantly the dynamics and severity of both effects.

## 4. Conclusion

The present study revealed the nonlinear acousto-hydrodynamics of standing surface acoustic wave (sSAW) interactions with a liquid film, its acoustic stabilization, patterning, and aerosol droplet generation. Utilizing very high temporal and spatial microscopic resolution allowed for a clear observation of the liquid atomization zone on the piezoelectric substrate. The fluid dynamics in resonant SAW aerosol generators were comprehensively elucidated, helping to understand the physics of the induced liquid patterns in the atomization zone, which supports cavitation alongside capillary wave excitation.

For the first time, the dynamics of the cyclic, spatiotemporal fluid dome formation under sSAW are investigated experimentally and supported by analytic models. The acoustics in the domes are modeled by balancing the capillary and acoustic pressures. We speculate that the domes become acoustic resonators with oscillating pressure amplitudes in which cavitation bubbles can be nucleated, grow, and collapse, as observed in the high-speed videos. Photographic evidence reveals that these bubbles are responsible for the ejection of specific droplet fractions from the liquid domes.

In addition, we found an analytic explanation for large-volume oscillations from the cavitation bubbles with the very high pressures created within the fluid domes under resonance. A simple spherical bubble model reveals that, although the bubbles are driven far above their resonance, through the nonlinearities, large bubble volume oscillations can be involved. We were able to quantitatively demonstrate that at a pressure of 100 bar, the bubble oscillations become exceedingly nonlinear, resulting in a greater compression rather than an expansion of the bubble, and eventually a violent collapse, possibly linked to droplet ejection.

Interestingly, the locations where cavitation bubble oscillations were observed are identical to the erosion seen on the substrate. This hints at cavitation-based erosion mechanics in aerosol generation SAW devices, which is of high significance for all fields of their application and may limit their applicability in the medical field.

The investigations in this study are contributing to a deeper understanding of sSAW atomization, therefore enhancing the predictability and practicality of such devices in real-world applications. Further research is needed to understand the cause of bubble nucleation and the limitation of cavitation erosion, as well as the dynamics of acoustowetting involved.

## Acknowledgements

This work was supported by the WIPANO project "MehrZAD" and by the German Research Foundation (DFG-ANR Grant AERONEMS 53301) (MR). J.M.R acknowledges support from the Alexander von Humboldt Foundation (Germany) through the Georg Forster Research Fellowship.